\documentclass[letterpaper, 10 pt, conference]{ieeeconf}  

\IEEEoverridecommandlockouts                              

\usepackage{url}
\usepackage{cite}
\usepackage{amsmath,amssymb,amsfonts}
\usepackage{algorithmic}
\usepackage{graphicx}
\usepackage{textcomp}
\usepackage[dvipsnames]{xcolor}
\usepackage{soul}
\usepackage{multirow}
\usepackage{tikz}
\usetikzlibrary{patterns,decorations.pathmorphing,positioning}

\def\BibTeX{{\rm B\kern-.05em{\sc i\kern-.025em b}\kern-.08em
    T\kern-.1667em\lower.7ex\hbox{E}\kern-.125emX}}

\newcommand{\vo}[1]{\boldsymbol{#1}}
\newcommand{\mo}[1]{\boldsymbol{#1}}
\newcommand{\A}{\vo{A}}
\newcommand{\B}{\vo{B}}

\newcommand{\x}{\vo{x}}
\newcommand{\y}{\vo{y}}

\newcommand{\w}{\vo{w}}
\newcommand{\n}{\vo{n}}

\newcommand{\s}{\vo{s}}

\newcommand{\norm}[2]{\left\lVert#1\right\rVert_{#2}}

\newcommand{\mub}{\vo{\mu}}

\newcommand{\xdot}{\dot{\vo{x}}}

\newcommand{\set}[1]{\mathcal{#1}}
\newcommand{\Exp}[1]{\mathbb{E}\left[#1\right]}
\newcommand{\I}[1]{\vo{I}_{#1}}

\newcommand{\K}{\vo{K}}
\newcommand{\W}{\vo{W}}

\newcommand{\R}{\vo{R}}
\newcommand{\M}{\vo{M}}
\newcommand{\N}{\vo{N}}
\newcommand{\Q}{\vo{Q}}

\renewcommand{\P}{\mo{P}} 
\newcommand{\Real}{\mathbb R}

\newcommand{\inner}[1]{\left\langle \vo{e}\phi_i\right\rangle}

\newcommand{\eqnlabel}[1]{\label{eqn:#1}}

\newcommand{\eqn}[1]{(\ref{eqn:#1})}

\newcommand{\fig}[1]{Fig. (\ref{fig:#1})}

\DeclareMathAlphabet{\mathbfsf}{\encodingdefault}{\sfdefault}{bx}{n}

\newcommand{\C}{\vo{C}}

\newcommand{\domain}[1]{\set{D}}

\newcommand{\diag}{\textbf{diag}}

\newcommand{\trace}[1]{\textbf{trace}\left( #1 \right)}

\title{\LARGE \bf Sparse Sensing Architectures with Optimal Precision for Tracking Multi-agent Systems in Sensing-denied Environments}
\author{Vedang M. Deshpande$^{1}$ and Raktim Bhattacharya$^{2}$
\thanks{This work was supported by the National Science Foundation (grant
number: 1762825).}
\thanks{$^{1}$Vedang M. Deshpande is a Ph.D. student in Aerospace Engineering, Texas A\&M University, College Station, TX 77843, USA. {\tt\small vedang.deshpande@tamu.edu}}%
\thanks{$^{2}$Raktim Bhattacharya is Associate Professor in Aerospace Engineering,
Electrical \& Computer Engineering, Texas A\&M University, College Station, TX 77843, USA. {\tt\small raktim@tamu.edu}}}

\begin{document}
\maketitle
\thispagestyle{empty} 

\begin{abstract}
In this paper the tracking problem of multi-agent systems, in a particular scenario where a segment of agents entering a sensing-denied environment or behaving as non-cooperative targets, is considered. The focus is on determining the optimal sensor precisions while simultaneously promoting sparseness in the sensor measurements to guarantee a specified estimation performance. The problem is formulated in the discrete-time centralized Kalman filtering framework. A semi-definite program subject to linear matrix inequalities is solved to minimize the trace of  precision matrix which is defined to be the inverse of sensor noise covariance matrix. Simulation results expose a trade-off between sensor precisions and sensing frequency.
\end{abstract}

\begin{keywords}
GPS-denied, sensor precision, sparse sensing, convex optimization.
\end{keywords}

\section{Introduction}
\begin{figure}[htb]
    \centering
    \includegraphics[trim=0cm 0cm 0cm 0cm,clip,width=0.32\textwidth]{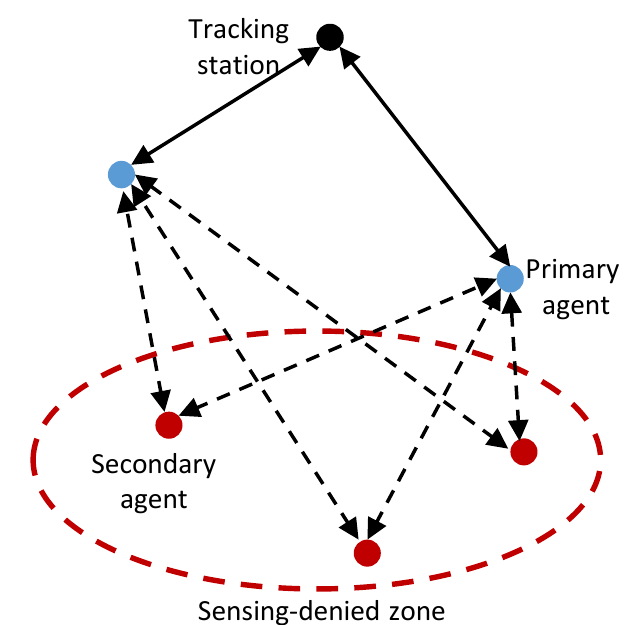}
    \caption{Multi-agent system in a sensing-denied zone.}
    \label{fig:big_picture}
\end{figure}

We consider the tracking problem of multi-agent systems. In particular, we are interested in a scenario wherein some agents can not be tracked using the tracking station, referred herein as \textit{secondary} agents. We refer the agents which can be directly tracked as \textit{primary} agents. We assume that the secondary agents can be tracked using the sensors onboard primary agents. See \fig{big_picture} as an illustration of this.

The primary agents, shown by blue circles in \fig{big_picture}, can be tracked from the primary sensor or the tracking station (black circle). The ellipse shown by red dotted line denotes the sensing-denied zone in which the secondary agents (red circles) can not be tracked directly from the tracking station. However, sensors aboard primary agents are used to track the secondary agents.

 \fig{big_picture} illustrates many practical applications of interest. For example, consider a  scenario in which emergency responders (secondary agents) are in a GPS-denied environment. Whereas, ground or air vehicles (primary agents) are able to track the responders using vision-based sensors. As an another example, consider exploration or surveillance mission of a sensing-denied landscape like tunnel, mine or a crater on a planet, and the primary agents are equipped with ranging equipment to track the secondary agents.

There exists substantial literature on control, estimation and navigation in sensing-denied, particularly, GPS-denied environments e.g. see \cite{demoz2016tac, demoz2014nav, gpsDenied2014Hamid, gpsDenied2011RANGE, gpsDenied2020peer2peer, gpsDenied2007borensteinBoot, gpsDenied2013saripalli, gpsDenied2017swarmUAV, gpsDenied2015tac} and the references therein.
In these works, agents \textit{actively} estimate their own states using various onboard sensors such as inertial measurement units \cite{demoz2014nav, gpsDenied2007borensteinBoot, gpsDenied2020peer2peer}, laser or radio based ranging instruments \cite{gpsDenied2011RANGE, gpsDenied2020peer2peer}, or vision based systems \cite{ gpsDenied2013saripalli, gpsDenied2015tac, gpsDenied2017swarmUAV}.

The problem under consideration in this paper (\fig{big_picture}) is slightly different as we are interested in \textit{passive} tracking of the secondary agents using sensors aboard primary agents. This is an extensible scenario capable of accommodating non-cooperative targets, e.g. space objects or enemy vehicles \cite{noncoop2006space, das2019laserOT, noncoop2017space, noncoop2016dong}, as secondary agents.

While tracking a multi-agent system, it is crucial that the estimation errors are bounded, and the number of measurements (or sensors) required to achieve a certain performance is minimal. In general, higher sensor precisions imply higher economic costs, thus, it is also desirable that the sensors require minimal precisions.
Therefore, the objective is to design a sparse sensing architecture with minimal sensor precisions for the system shown in \fig{big_picture} such that the estimation errors are within the specified performance bounds.

Sparse sensor selection \cite{Sundaram2017automatica, pappas2016acc, Wolfrum2014tac, Jovanovic2019TAC} and sensor scheduling \cite{scheduling2006he, scheduling2015jawaidAuto, scheduling2017hanAuto, scheduling2013shiSCL} are well-studied problems. However, only few formulations \cite{skelton2008jour, deshpande2021cdcLCSS, das2020mrKalman} allow
simultaneous minimization of sensor precisions while designing sparsity promoting sensing architectures.
Therefore, to meet the aforementioned objective, we employ the formulation discussed in \cite{das2020mrKalman}, which achieves a sparse sensing configuration with optimal precisions by minimizing of $l_1$-norm of the precision vector in the discrete-time Kalman filtering framework.

The rest of the paper is organized as follows. Section \ref{sec:prob} formulates the multi-agent tracking problem in the Kalman filtering framework. Numerical simulation results for a test example involving three agents are shown in Section \ref{sec:sim} followed by concluding remarks and future research directions in Section \ref{sec:concl}.

\section{Problem Formulation} \label{sec:prob}
\subsection{Notation}
The set of real numbers is denoted by $\Real$. Bold uppercase (lowercase) letters denote matrices (column vectors).
$\I{}$ and $\vo{0}$ respectively denote an identity matrix and a zero matrix of suitable dimensions.
$\P^T$ denotes transpose of $\P$. We use the notation  ($\P\geq0$) $\P>0$ to denote symmetric positive (semi-)definite matrices. Analogous notations are used to denote negative (semi-)definite matrices.
$\diag(\x)$ denotes a diagonal matrix with diagonal elements as the vector $\x$. Similarly, $\diag\left(\P_1,\P_2,\cdots,\P_N \right)$ denotes a block diagonal matrix. All powers and inequalities involving  vectors are to be interpreted elementwise. $\Exp{\cdot}$ denotes the expectation operator, and $\delta_{ij}$ denotes the Kronecker delta defined as $\delta_{ij}:=1$ if $i=j$, and $\delta_{ij}:= 0$ if $i\neq j$.

\subsection{System equations}
Let us assume that there are $N_1$ \textit{primary} agents that can be directly tracked from the tracking stations. Let there be $N_2$ \textit{secondary} agents that are in the sensing-denied zone where tracking stations are incapable of tracking them.
The dynamic models for the primary and secondary agents are given by
\begin{align*}
  \x^{(i)}_{k+1} &= \A^{(i)}_k\x^{(i)}_k  + \B^{(i)}_k\w^{(i)}_k, \ \ i = 1,\cdots, (N_1+N_2),
\end{align*}
where $k = 0, 1 \cdots$ denotes the temporal index, and $i = 1,\cdots, N_1$ denotes dynamics of the primary agents, and $i = N_1+1,\cdots,N_1+N_2$ denotes  dynamics of the secondary agents. $\x^{(i)}_k$ denotes the state vector of the $i^{\text{th}}$ agent. The process noise encountered by the $i^{\text{th}}$ agent is denoted by $\w^{(i)}_k$ and it is assumed to be zero-mean Gaussian noise process with covariance
$$\mathbb{E}[{\w^{(i)}_k\w^{(i)}_l}] = \Q_k^{(i)} \delta_{kl}.$$
The real time-varying system matrices $\A^{(i)}_k$, $\B^{(i)}_k$ are of appropriate dimensions. We define an augmented system as follows
\begin{align}
  \x_{k+1} &= \A_k\x_k  + \B_k\w_k, \eqnlabel{sys_proc}
\end{align}
\begin{align*}
  \x_k &:= \begin{bmatrix} \x^{(1)}_{k} \\ \vdots \\ \x^{(N_1+N_2)}_{k}  \end{bmatrix}, \ \   \w_k:= \begin{bmatrix} \w^{(1)}_{k} \\ \vdots \\ \w^{(N_1+N_2)}_{k} \end{bmatrix}, \\
  \A_k &:= \diag\left(\A^{(1)}_k, \cdots, \A^{(N_1+N_2)}_k\right) \\
  \B_k &:= \diag\left(\B^{(1)}_k, \cdots, \B^{(N_1+N_2)}_k\right),
\end{align*}
and $\w_k$ is also a zero-mean Gaussian noise process with augmented covariance matrix
$$\Q_k : = \diag\left(\Q^{(1)}_k, \cdots, \Q^{(N_1+N_2)}_k\right).$$
As mentioned earlier, tracking stations can track the primary $N_1$ agents, and the secondary agents are tracked indirectly using sensors aboard the primary agents. All measurements $\y_k$ are written as the following measurement equation in terms of the augmented state $\x_k$
\begin{align}
  \y_k = \C_k\x_k + \n_k, \eqnlabel{sys_meas}
\end{align}
where $\C_k$ is the time-varying output matrix, and $\n_k$ is the zero-mean Gaussian sensor noise with covariance
 $$\mathbb{E}[{\n_k\n_l}] = \R_k \delta_{kl},$$
 where $\R_k$ is assumed to be a diagonal matrix.

The initial condition $\mub_0:=\Exp{\x_0}$, $\vo{\Sigma}_0:=\Exp{(\x_0-\mub_0)(\x_0-\mub_0)^T}$ are assumed  to be known, and the initial state variable $\x_0$, process noise $\w_k$, and sensor noise $\n_k$ are assumed to be mutually independent.

For a system given by \eqn{sys_proc} and \eqn{sys_meas}, the sequential optimal Kalman filter takes the following form \cite{junkins2004book}
\begin{align*}
  \text{Prior mean: } \mub_k^{-} &= \A_k\mub_{k-1}^{+}, \\
  \text{Prior covariance: } \vo{\Sigma}_k^{-} &= \A_k \vo{\Sigma}_{k-1}^{+}\A_k^T + \B_k\Q_k\B_k^T, \\
  \text{Kalman gain: } \K_k &= \vo{\Sigma}_k^{-}\C_k^T\left[\C_k\vo{\Sigma}_k^{-}\C_k^T + \R_k\right]^{-1}, \\
  \text{Posterior mean: } \mub_k^{+} &= \mub_k^{-} + \K_k(\y_k - \C_k\mub_k^{-}) ,\\
  \text{Posterior covariance: } \vo{\Sigma}_k^{+} &= (\I{} - \K_k\C_k)\vo{\Sigma}_k^{-},
\end{align*}
with $\mub_0^{+} = \mub_0$ and $\vo{\Sigma}_0^{+} = \vo{\Sigma}_0$. For a given $\R_k$, the above equations provide an estimate of the state with the least error variance. Herein, we treat $\R_k$ as a variable, as discussed next.

 The precision of a sensor channel is defined as inverse of the signal variance. Let us define $\vo{S}_k:= \diag (\s_k) := \R_k^{-1}$. Therefore, $\vo{S}_k$ is interpreted as the precision matrix. As discussed earlier, in this work we are interested in determining a sparse sensor configuration, with the minimum required precision. This can be achieved by minimizing $\textbf{trace}(\vo{S}_k)$ or $\norm{\s_k}{1}$, since minimization of $l_1$-norm promotes sparsity \cite{boyd2008weightedL1}.

 If precision of a sensor channel is zero, then that sensor has infinite noise variance, hence it is removed from the sensor architecture. While identifying a sparse sensor configuration, we require that the posterior covariance should satisfy $\trace{\vo{\Sigma}_k^{+}} \leq \gamma$, for a specified $\gamma > 0$, so that the estimation errors are bounded.

We assume that the update step in the Kalman filter is carried out every $p$ time steps, thus, measurements over  a finite horizon of $p$ time steps are accumulated and used in a batch processing framework, which is discussed next.

\subsection{Batch processing}
Our objective is to determine a sparse sensing configuration with minimum sensor precision, given posterior statistics (mean and covariance) at time step $kp$, and the measurements over the time steps from $kp+1$ to $(k+1)p$. To this end, we write an augmented system as follows
\begin{align*}
  \vo{\bar{x}}_k &:= \begin{bmatrix} \x_{kp+1}\\ \vdots \\ \x_{(k+1)p} \end{bmatrix}, \
  \vo{\bar{w}}_k := \begin{bmatrix} \w_{kp}\\ \vdots \\ \w_{(k+1)p-1} \end{bmatrix}, \\
  \vo{\bar{y}}_k &:= \begin{bmatrix} \y_{kp+1}\\ \vdots \\ \y_{(k+1)p} \end{bmatrix}, \
  \vo{\bar{n}}_k := \begin{bmatrix} \n_{kp+1}\\ \vdots \\ \n_{(k+1)p} \end{bmatrix},
\end{align*}
\begin{align*}
  \vo{\bar{x}}_k  &= \vo{\bar{A}}_k\x_{kp} + \vo{\bar{B}}_k  \vo{\bar{w}}_k, \\
  \vo{\bar{y}}_k  &= \vo{\bar{C}}_k\vo{\bar{x}}_k + \vo{\bar{n}}_k,
\end{align*}
where,
\begin{align*}
  \vo{\bar{A}}_k &:= \begin{bmatrix} \A_{kp} \\ \A_{kp+1}\A_{kp} \\ \prod_{i=0}^{p-1}\A_{kp+i} \end{bmatrix},  \\
  \vo{\bar{B}}_k &:= \begin{bmatrix} \B_{kp} & \vo{0} & \cdots & \vo{0} \\
                                    \A{kp+1}\B_{kp} & \B_{kp+1} & \cdots & \vo{0} \\
                                    \vdots & \vdots & \ddots & \vdots \\
                                    \prod_{i=1}^{p-1}\A_{kp+i}\B_{kp} & \cdots & \cdots & \B_{(k+1)p-1}
                      \end{bmatrix} \\
  \vo{\bar{C}}_k &:= \diag\left(\C_{kp+1},\cdots,\C_{(k+1)p}\right).
\end{align*}
Note that the overhead bar indicates an augmented variable.
The covariances of the augmented process noise $\vo{\bar{w}}_k$ and sensor noise $\vo{\bar{n}}_k$ are given respectively as
\begin{align*}
  \vo{\bar{Q}}_k &:= \diag\left(\Q_{kp}, \cdots, \Q_{(k+1)p-1}\right) \\
  \vo{\bar{R}}_k &:= \diag\left(\R_{kp+1}, \cdots, \R_{(k+1)p}\right).
\end{align*}
The propagated prior statistics of the augmented state $\vo{\bar{x}}_k$, i.e. $\vo{\bar{\mu}}_k$ and $\vo{\bar{\Sigma}}_k$, are written in terms of the posterior statistics of $\x_{kp}$ as
\begin{align*}
  \vo{\bar{\mu}}_k^{-} &:= \Exp{\vo{\bar{x}}_k} = \vo{\bar{A}}_k \mub_{kp}^{+} \\
  \vo{\bar{\Sigma}}_k^{-} &:= \Exp{ (\vo{\bar{x}}_k-\vo{\bar{\mu}}_k)(\vo{\bar{x}}_k-\vo{\bar{\mu}}_k)^T  }  \\
 &= \vo{\bar{A}}_k \vo{\Sigma}_{kp}^{+} \vo{\bar{A}}_k ^T + \vo{\bar{B}}_k \vo{\bar{Q}}_k \vo{\bar{B}}_k ^T.
\end{align*}
Similarly, the updated posterior statistics of the augmented state are obtained using the standard Kalman update equation as follows.
\begin{align*}
    \vo{\bar{\mu}}_k^{+}  &= \vo{\bar{A}}_k \mub_{kp}^{+}  + \bar{\vo{K}}_k(\vo{\bar{y}}_k -\vo{\bar{C}}_k \vo{\bar{A}}_k \mub_{kp}^{+} ) ,\\
    \vo{\bar{\Sigma}}_k^{+} &= (\I{} - \bar{\vo{K}}_k\vo{\bar{C}}_k)\vo{\bar{\Sigma}}_k^{-}(\I{} - \bar{\vo{K}}_k\vo{\bar{C}}_k)^T + \bar{\vo{K}}_k\vo{\bar{R}}_k\bar{\vo{K}}_k^T,
\end{align*}
where $\bar{\vo{K}}_k$ is the Kalman gain for the augmented system. The optimal gain which minimizes the $\trace{  \vo{\bar{\Sigma}}_k^{+}}$ is given by $\bar{\vo{K}}_k = \vo{\bar{\Sigma}}_k^{-} \vo{\bar{C}}_k^T\left[\vo{\bar{C}}_k\vo{\bar{\Sigma}}_k^{-} \vo{\bar{C}}_k^T + \vo{\bar{R}}_k\right]^{-1}$. However, herein $\vo{\bar{R}}_k$ and hence $\bar{\vo{K}}_k$ are variables.

We require that the inequality $\trace{\vo{\Sigma}_{(k+1)p}^{+}} \leq \gamma$ is satisfied. Therefore, let us define a masking matrix $\M:=[\vo{0}\ \ \I{}]$ of appropriate dimensions such that we have
\begin{align*}
  \mub_{(k+1)p}^{+} = \M\vo{\bar{\mu}}_k^{+}, \; \text{ and } \; \vo{\Sigma}_{(k+1)p}^{+} = \M\vo{\bar{\Sigma}}_k^{+}\M^T.
\end{align*}
Then the inequality  $\trace{\vo{\Sigma}_{(k+1)p}^{+}} \leq \gamma$, or $\trace{\M\vo{\bar{\Sigma}}_k^{+}\M^T} \leq \gamma$ is equivalently written as
\begin{align*}
  \trace{\W} \leq \gamma \\
  \W - \N\vo{\bar{\Sigma}}_k^{-}\N^T - \M\bar{\vo{K}}_k\vo{\bar{R}}_k\bar{\vo{K}}_k^T\M^T \geq 0
\end{align*}
where $\W>0$ and $\N:=\M(\I{}-\bar{\vo{K}}_k\vo{\bar{C}}_k)$. Using Schur complement lemma, we get the following LMI
\begin{align}
\begin{bmatrix}
  \W & \M(\I{}-\bar{\vo{K}}_k\vo{\bar{C}}_k)\sqrt{\vo{\bar{\Sigma}}_k^{-}} & \M\bar{\vo{K}}_k \\
  \ast & \I{} & \vo{0} \\
  \ast & \ast & \diag(\vo{\bar{s}}_k)
\end{bmatrix} \geq 0 \eqnlabel{lmi}
\end{align}
where $\sqrt{\vo{\bar{\Sigma}}_k^{-}}$ is the principal matrix square root of ${\vo{\bar{\Sigma}}_k^{-}}$, and we substitute $\vo{\bar{R}}_k^{-1} = \vo{\bar{S}}_k = \diag(\vo{\bar{s}}_k)$ \cite{das2020mrKalman}. Then the optimal precision vector $\vo{\bar{s}}_k^{\ast}$ is determined by minimization of the $l_1$-norm or in a more general setting, weighted $l_1$-norm of $\vo{\bar{s}}_k$. Weighted $l_1$-norm is defined as
$$\norm{\vo{\bar{s}}_k}{1,\vo{\rho}} := \vo{\rho}^T\vo{\bar{s}}_k,$$
where $\vo{\rho}>0$ is a specified weight vector of the same dimensions as $\vo{\bar{s}}_k$.

The precisions of sensors are generally upper bounded due to physical constraints, i.e.
\begin{align}
  0\leq\vo{\bar{s}}_k\leq\vo{\bar{s}}_{\max}. \eqnlabel{upperbound}
\end{align}
Therefore, the optimal $\vo{\bar{s}}_k^{\ast}$  is given as
\begin{align}
  \vo{\bar{s}}_k^{\ast} := \arg \left\{\min \norm{\vo{\bar{s}}_k}{1,\vo{\rho}} \text{ subject to \eqn{lmi}, \eqn{upperbound}} \right\}. \eqnlabel{optim}
\end{align}
The sparseness of the solution $\vo{\bar{s}}_k^{\ast}$ can be improved by implementing iterative reweighting schemes such as \cite{boyd2008weightedL1}. The problem \eqn{optim} is solved iteratively with weights $\vo{\rho}^{(j+1)} = (\vo{\bar{s}}_k^{(j)\ast} + \epsilon\vo{1})^{-1}$, where $j$ is the iteration index, $\vo{1}$ denotes a column vector of all ones and the inverse is interpreted elementwise, $\vo{\bar{s}}_k^{(j)\ast}$ is the solution determined in the $j^{\text{th}}$ iteration, and $\epsilon>0$ is a small number which ensures that the weights are well-defined.

A scenario in which some sensors can not be used, e.g. primary agents are obscured by some obstacles, can be easily accounted for in the optimization problem \eqn{optim} by simply imposing a linear constraint to enforce the corresponding elements of $\vo{\bar{s}}_k$ to be zero.

\section{Simulation Results} \label{sec:sim}
\begin{figure}[htb]
    \centering
    \includegraphics[trim=0cm 0.1cm 0cm 0cm,clip,width=0.38\textwidth]{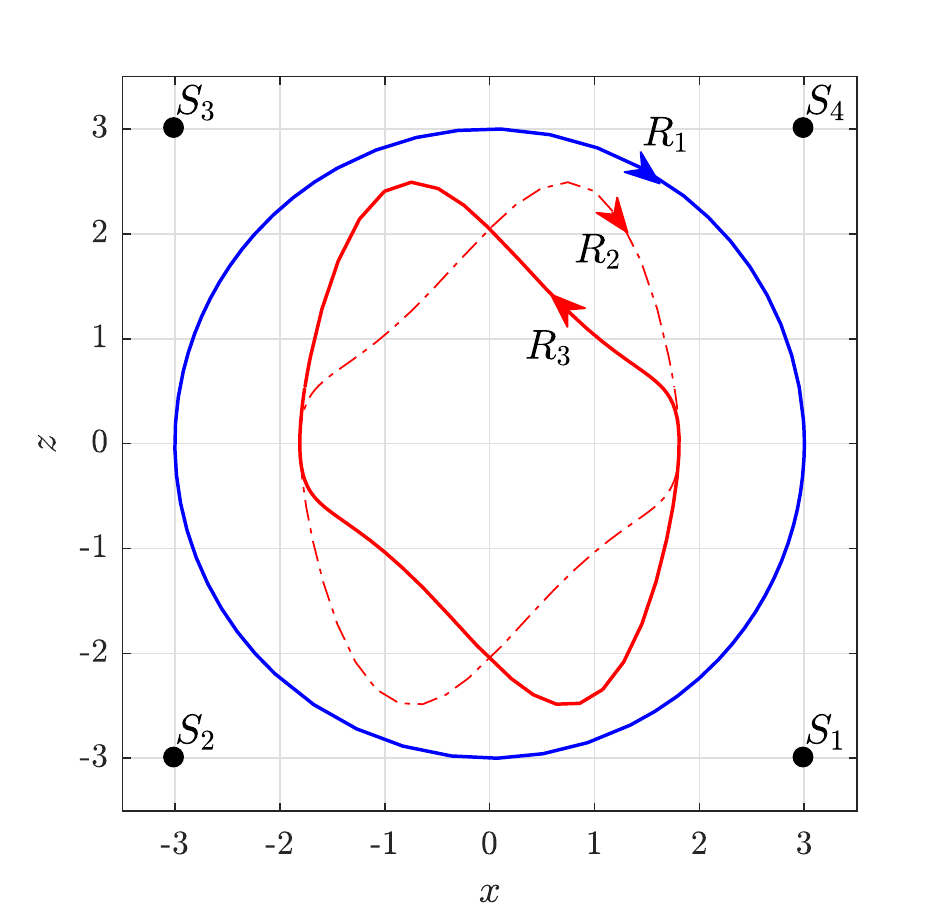}
    \caption{Tracking stations $S_i$, $i=1,2,3,4$, and agents $R_i$, $i=1,2,3$. $R_2$ and $R_3$ are secondary agents in the sensing-denied zone.}
    \label{fig:setup}
\end{figure}

 \fig{setup} shows the nominal planar trajectories of three different agents.
The primary agent $R_1$ shown in blue, moves in a circular trajectory. The stationary tracking stations $S_i$, $i = 1,2,3,4$ can track only $R_1$. We assume that the position coordinates of tracking stations are known, and they can measure the range of $R_1$ from their respective locations. The secondary agents $R_2$ and $R_3$ (shown by red periodic trajectories) are in sensing-denied zone, i.e. the tracking stations $S_i$ can not directly measure the ranges of $R_2$ and $R_3$. Instead, the primary agent $R_1$ can measure the relative ranges of $R_2$ and $R_3$ from its location.

Let $(x_i,z_i)$ denote the position coordinates of the $i^{\text{th}}$ agent $R_i$. The nonlinear equations of the  trajectories are given by
\begin{equation}
  \begin{aligned}
    \dot{x}_1 &= z_1, \ \ \dot{z}_1 = -x_1 + w_1,\\ 
    \dot{x}_2 &= z_2, \ \ \dot{z}_2 = (1-x_2^2/c^2)z_2 - x_2/c + w_2, \\ 
    \dot{x}_3 &= -z_3, \ \ \dot{z}_3 = (1-x_3^2/c^2)z_3 - x_3/c + w_3, 
  \end{aligned}  \eqnlabel{traj_dyn}
\end{equation}
where $w_i$ denotes the zero-mean Gaussian process noise encountered by the $i^{\text{th}}$ agent, and $c=0.9$. Let us define
$$\x:=\begin{bmatrix} {x}_1 & {z}_1 &  {x}_2 & {z}_2 &  {x}_3 & {z}_3 \end{bmatrix}^T.$$

The state variable $\x$ is decomposed into a nominal variable $\vo{\hat{x}}$ and the perturbation $\vo{\tilde{x}}$ such that
\begin{align}
  \x(t) = \vo{\hat{x}}(t) + \vo{\tilde{x}}(t). \eqnlabel{decompose_x}
\end{align}
The nominal trajectories $\vo{\hat{x}}(t)$ are shown in \fig{setup} for the initial condition
$$\vo{\hat{x}}(0)=\begin{bmatrix} 3 & 0 & 1.7636 & 0.5215 & -1.7636 & 0.5215 \end{bmatrix}^T.$$

We linearize the dynamics \eqn{traj_dyn} about the nominal trajectory $\vo{\hat{x}}(t)$ to get a linear continuous-time periodic system as follows.
\begin{align}
  \vo{\dot{\tilde{x}}} = \A_c(t)\vo{\tilde{x}}(t) + \B_c\w(t), \eqnlabel{linearized_sys_CT}
\end{align}
where  $\w(t):= \left[w_1 \,\, w_2\,\, w_3\right]^T$ denotes the process noise. Subscript $c$ indicates that system matrices are continuous-time. $\A_c(t)\in\Real^{6\times 6}$ is the Jacobian of $\xdot(t)$ with respect to $\x(t)$ evaluated at $\vo{\hat{x}}(t)$, and $\B_c = \I{} \otimes [0 \; 1]^T \in \Real^{6\times 3}$.

\begin{figure}[htb]
    \centering
    \includegraphics[trim=0cm 0.5cm 0cm 1cm,clip,width=0.5\textwidth]{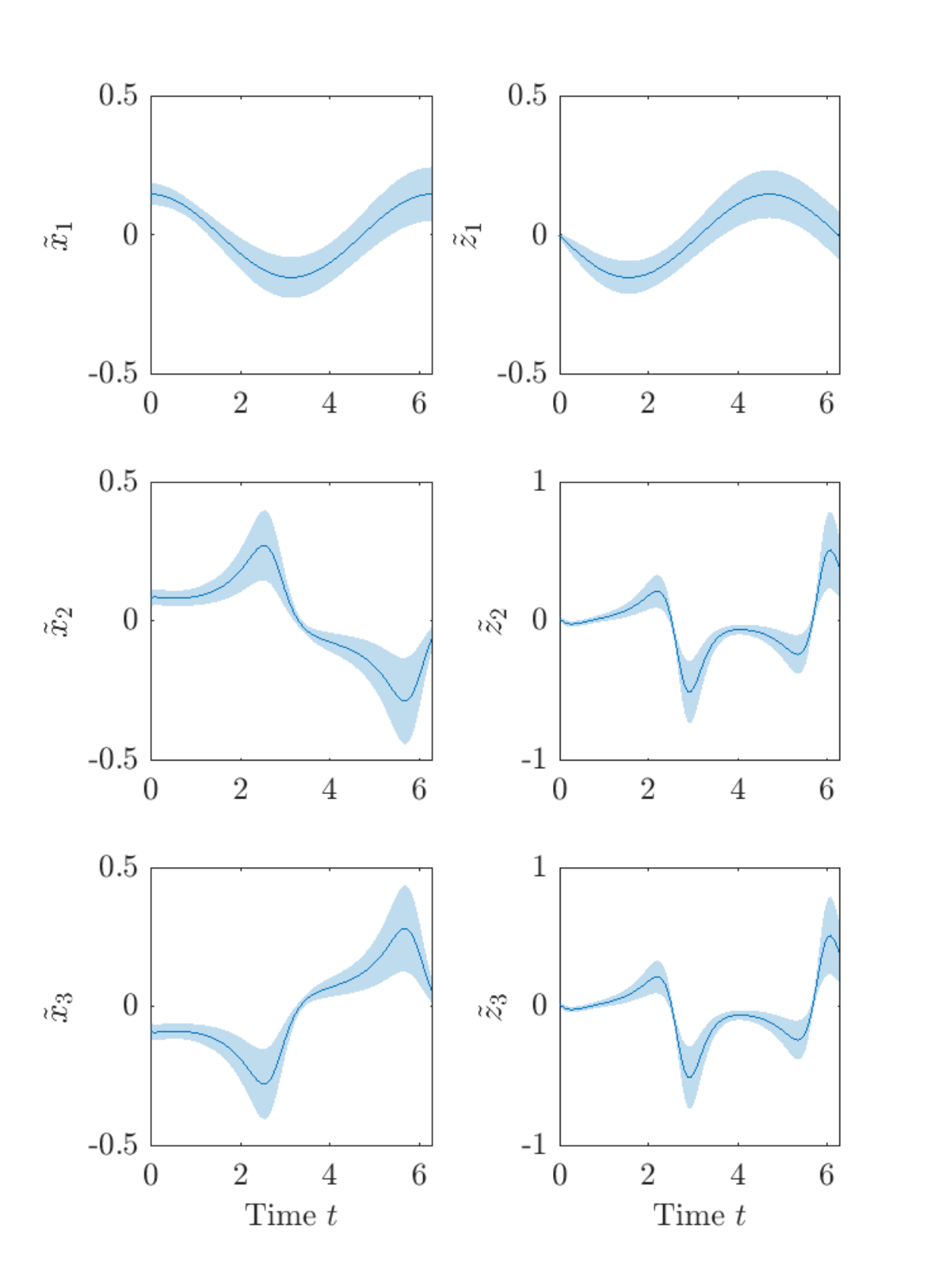}
    \caption{Mean and covariance evolution for linearized system \eqn{linearized_sys_CT} obtained using \eqn{mean_cov_CT}. The perturbation mean is shown by the solid line, and the shaded region highlights the $1\sigma$-bound. }
    \label{fig:mean_cov}
\end{figure}

The quantity of interest $\x(t)$ is a stochastic variable since the perturbation $\vo{\tilde{x}}(t)$ is stochastic  and governed by \eqn{linearized_sys_CT}. Let us denote mean and covariance of $\vo{\tilde{x}}(t)$ as $\mub(t):=\Exp{\vo{\tilde{x}}(t)}$ and $\vo{\Sigma}(t):=\Exp{(\vo{\tilde{x}}(t) - \mub(t))(\vo{\tilde{x}}(t) - \mub(t))^T}$.
The mean and covariance evolution equations follow from  \eqn{linearized_sys_CT} as
\begin{subequations}
\begin{align}
  \dot{\vo{\mu}}(t) &= \A_c(t)\mub(t), \\
  \dot{\vo{\Sigma}}(t) &= \A_c(t)\vo{\Sigma}(t) + \vo{\Sigma}(t)\A_c^T(t) + \B_c\Q\B_c^T, \eqnlabel{cont_cov}
\end{align}\eqnlabel{mean_cov_CT}
\end{subequations}
where $\Q:=0.05^2\I{}\in\Real^{3\times 3}$ denotes the spectral density matrix of $\w(t)$. Initial statistics are assumed to be $\mub(0) = 0.05\ \vo{\hat{x}}(0)$ and $\vo{\Sigma}(0) = 0.1^2 \ \diag(|\mub(0)|)$. The temporal evolution of the mean and covariance obtained using \eqn{mean_cov_CT} is shown in \fig{mean_cov}.

The time period of the trajectories shown in \fig{setup} is $T_p=2\pi$. We assume that the range measurements are obtained 10 times in a period at time instants $t = t_k:= k\Delta t$, with $\Delta t = 0.1T_p$.
The nonlinear range measurements in terms of $(x_i,z_i)$ and the known positions of the tracking stations at $t = t_k$ are given as
\begin{align}
  \y_k := \begin{bmatrix} (x_1-3)^2 + (z_1+3)^2 \\ (x_1+3)^2 + (z_1+3)^2 \\  (x_1+3)^2 + (z_1-3)^2 \\  (x_1-3)^2 + (z_1-3)^2 \\ (x_2-x_1)^2 + (z_2-z_1)^2  \\ (x_3-x_1)^2 + (z_3-z_1)^2 \end{bmatrix}^{0.5}_{|t = t_k} + \n_k, \eqnlabel{yrange}
\end{align}
where $\n_k\in\Real^{6}$ is the measurement noise.
Peer-to-peer measurements \cite{gpsDenied2020peer2peer} e.g. relative range between $R_2$ and $R_3$, if available, can be easily augmented in \eqn{yrange}.

We discretize the linearized equation \eqn{linearized_sys_CT} in time interval $[0,T_p]$ with $\Delta t = 0.1T_p$. The discretized dynamics is given by
\begin{align}
  \vo{\tilde{x}}_{k+1}= \A_k\vo{\tilde{x}}_k + \B_k\w_k, \eqnlabel{disc_proc}
\end{align}
where  $\vo{\tilde{x}}_{k}:=\vo{\tilde{x}}(t_k)$, $\A_k:=\vo{\Phi}(t_{k+1},t_k)$, $\B_k = \I{}$,
$$\w_k = \int_{t_k}^{t_{k+1}}\vo{\Phi}(\tau,t_k)\B_c\w(\tau)d\tau,$$
and $\vo{\Phi}(\cdot,\cdot)$ denotes the state transition matrix. The discrete time process $\w_k$ has zero mean i.e. $\Exp{\w_k} = 0$ since $\Exp{\w(t)} = 0$. The covariance of $\w_k$, $\Q_k = \Exp{\w_k\w_k^T}$ is determined via $\vo{\Sigma}(t_{k+1})$ calculated using the continuous-time covariance propagation equation \eqn{cont_cov}, and the discrete-time covariance propagation  as
\begin{align*}
  \Q_k = \vo{\Sigma}(t_{k+1}) - \A_k\vo{\Sigma}(t_{k})\A_k^T.
\end{align*}

Similar to \eqn{decompose_x} and \eqn{linearized_sys_CT}, the linearized range measurement equation follows from \eqn{yrange},
\begin{align}
  \y_k = \vo{\hat{y}}_k + \vo{\tilde{y}}_k, \ \ \
  \vo{\tilde{y}}_k = \C_k\vo{\tilde{x}}_k + \n_k, \eqnlabel{disc_meas}
\end{align}
where $\C_k\in\Real^{6\times 6}$ is the Jacobian of $\vo{y}_k$ with respect to $\x(t)$ evaluated at $\vo{\hat{x}}(t_k)$.

With equations \eqn{disc_proc} and \eqn{disc_meas}, we have formulated the tracking problem in the form given by \eqn{sys_proc} and \eqn{sys_meas}, and we can solve the optimization problem \eqn{optim} for the system under consideration.  The performance requirement is specified as
$$\trace{\vo{\Sigma}^{+}(t_{10})} \leq 0.1 \ \trace{\vo{\Sigma}^{-}(t_{10})},$$
i.e. we require 90\% reduction in the trace of prior covariance matrix after update at the time step $k=10$.

\begin{figure}[htb]
    \centering
    \includegraphics[trim=0cm 1cm 0cm 0cm,clip,width=0.45\textwidth]{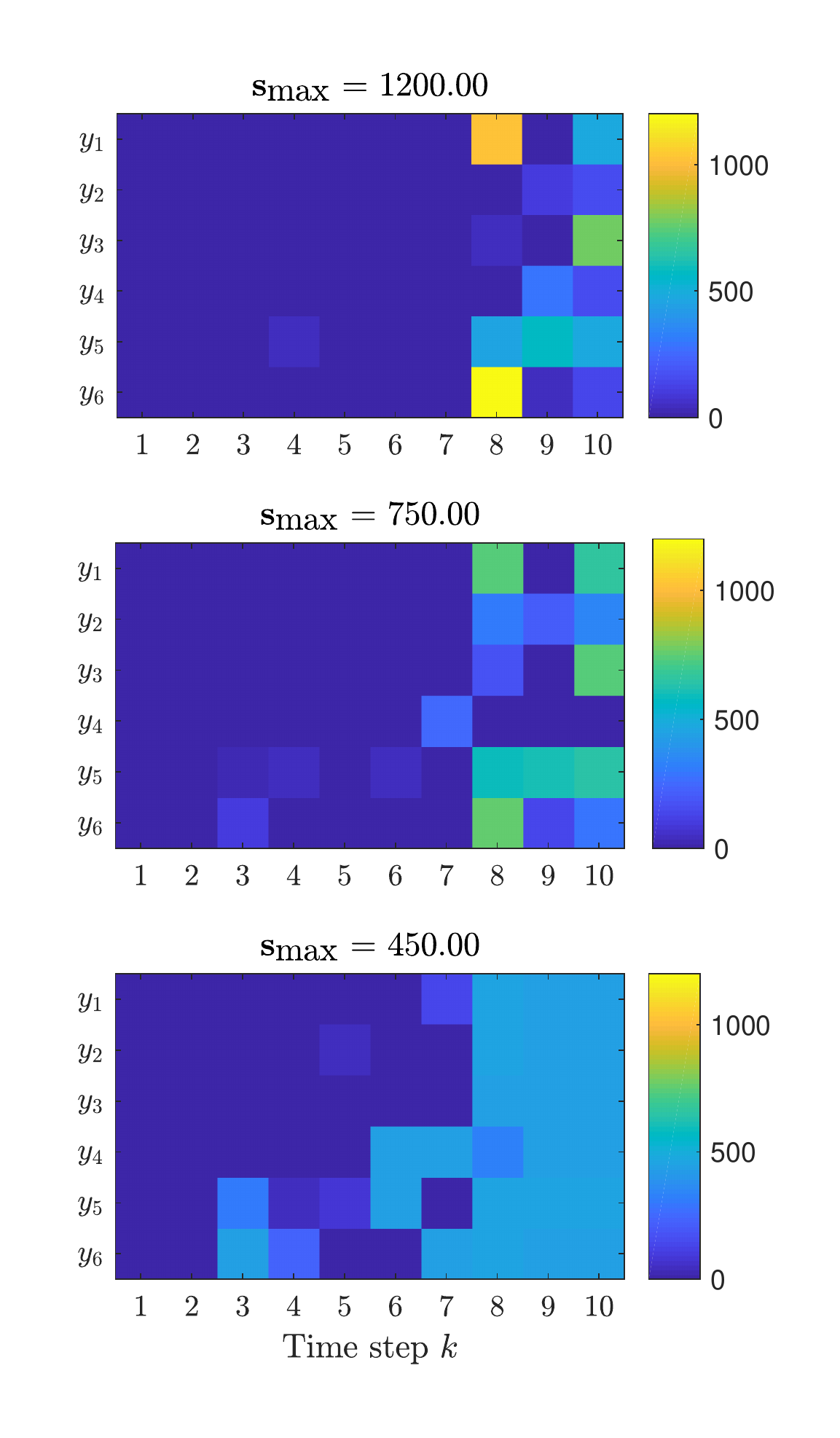}
    \caption{Sensing precisions (shown by colorbar) for different range measurements from time step $k=1$ to $k=10$.}
    \label{fig:heat}
\end{figure}

The optimal solution of \eqn{optim} is obtained using the solver \texttt{MOSEK}\cite{mosek} with \texttt{CVX} \cite{cvx} as a parser for three different values of $\vo{\bar{s}}_{\max}$, and shown in \fig{heat}. The blocks with dark blue color correspond to very low or zero precisions, indicating that those measurements are not required.

By solving \eqn{optim}, we have introduced sparseness in the sensing configuration. Since $\vo{\bar{s}}_k$ is a stacked vector of all sensor precisions at all time steps, the sparseness that we achieve is twofold. For example, see the top plot in \fig{heat} corresponding to $s_{\max} = 1200$.  At the time step $k=9$, only three measurements $y_2,y_4,y_5$ out of total available six measurements \eqn{yrange} are needed. We also observe that the most measurements are required for $k\geq 8$, thus introducing sensing sparseness temporally.

On decreasing the precision of sensors from $1200$ to $450$, we observe a reduction in the sparseness of measurements. In other words, the same filtering performance is achieved with fewer measurements by using sensors with higher precisions. This exposes a trade-off between sensor precision and sensing frequency.

\begin{figure}[htb]
    \centering
    \includegraphics[trim=0cm 0.25cm 0cm 0cm,clip,width=0.45\textwidth]{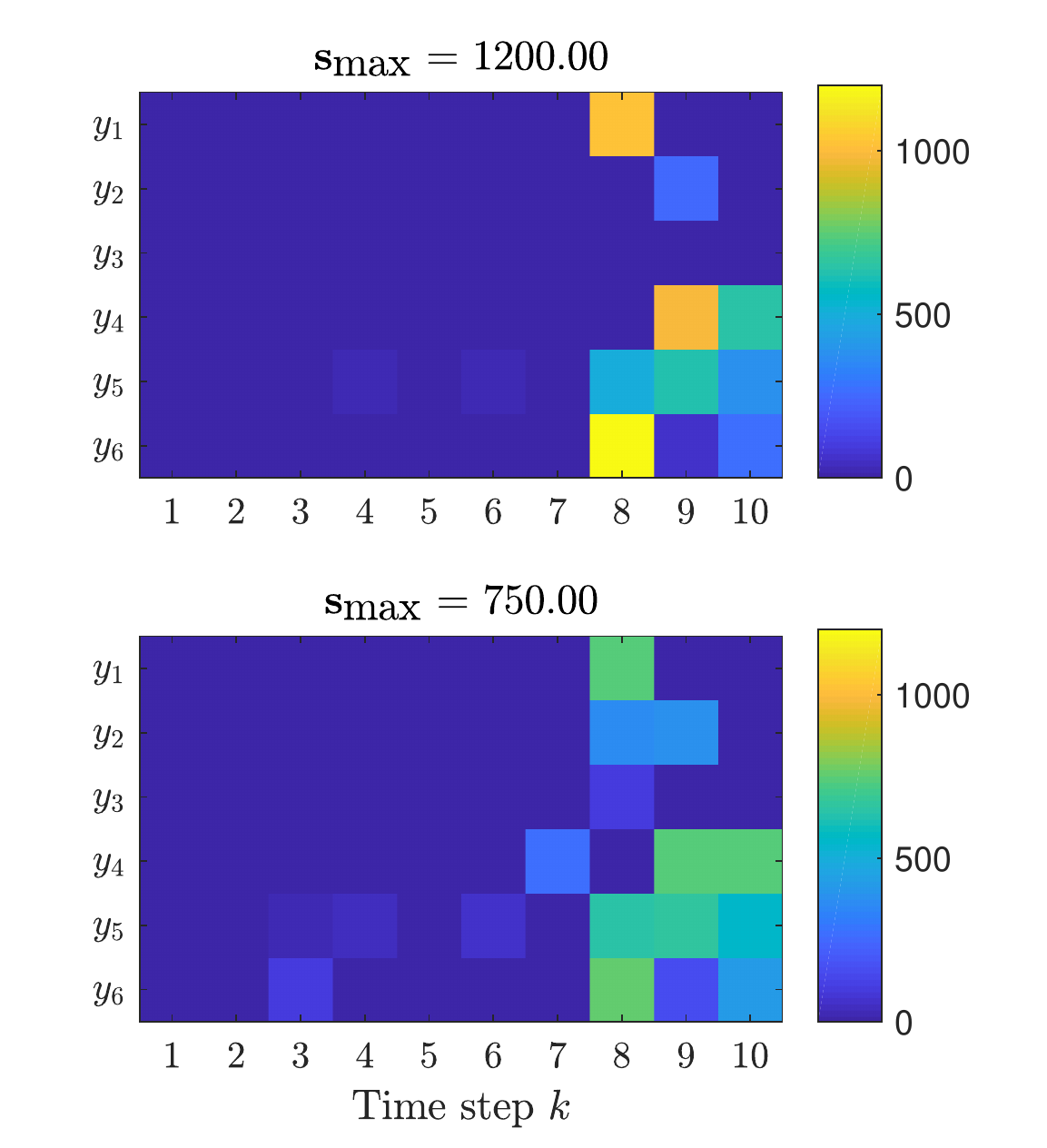}
    \caption{Sensing precisions (shown by colorbar) for different range measurements from time step $k=1$ to $k=10$. Measurements $y_1,y_2,y_3$ are unavailable at $k=10$.}
    \label{fig:heat2}
\end{figure}

Now, for the sake of argument, let us consider a case when the tracking stations $S_1$, $S_2$ and $S_3$ can not track the primary agent at time step $k=10$ due to some obstacles. This condition is incorporated in the optimization problem \eqn{optim} by constraining the precision values of $y_1,y_2,y_3$ at $k=10$ to zero, which is equivalent to not using those measurements in the Kalman update step. The optimal solution for this case is shown in \fig{heat2}.

Consider the plot for $s_{\max} = 750$ in \fig{heat2}. The unavailability of  $y_1,y_2,y_3$ at $k=10$ is compensated by the extra measurements of $y_4$ at $k=9,10$, which were not required in the previous case shown in \fig{heat}. We also observe that the required precisions for some measurements are higher in \fig{heat2} than \fig{heat}. Similar observation can be made for the plot corresponding to $s_{\max} = 1200$ as well.
The optimization problem is infeasible for $s_{\max} = 450$, and hence not shown in \fig{heat2}. It implies that the sensors with $s_{\max} = 450$ are not precise enough to guarantee the specified performance bound if the  measurements $y_1,y_2,y_3$ are unavailable at $k=10$.

\section{Conclusion} \label{sec:concl}
In this paper we considered the problem of tracking multi-agent systems in which some agents are non-cooperative targets or in a sensing-denied environment, and indirect measurements obtained by other agents are used to track the complete system. The objective of obtaining optimal sensor precisions while simultaneously promoting sparseness in the sensing architecture was achieved by formulating the problem in discrete-time Kalman filtering framework and minimizing $l_1$-norm of the precision vector. The optimization problem formulated as a semi-definite program (SDP) subject to linear matrix inequalities exposed a trade-off between sensor precisions and the number of measurements required to guarantee a certain estimation performance.

The dimension of the SDP grows quadratically with the number of agents in the system and the discrete-time horizon over which the problem is solved. Since general-purpose SDP solvers do not scale well with the increasing problem dimension, the development of customized solvers which exploit local problem structure is a topic of our ongoing research.
For the sake of simplicity, we did not consider communication or operational constraints on sensors, e.g. the maximum number of sensors that can simultaneously operate at a given instant. Our future work will incorporate constrained sensing and correlated sensor noise (i.e. non-diagonal covariance matrix) in the formulation.

\section*{Acknowledgment}
We want to thank Dr. Demoz Gebre-Egziabher from the University of Minnesota for introducing this problem to us and providing invaluable insights into this problem.



\bibliographystyle{unsrt}
\bibliography{root}
\end{document}